\documentclass[prb,showpacs]{revtex4}

\usepackage{amsmath,amssymb}

\usepackage{graphicx}

\begin{document}

\widetext

\title{Theory of the singlet exciton yield in light-emitting polymers}

\author{William Barford}

\affiliation{$^*$Department of Physics and Astronomy, University
of Sheffield, Sheffield, S3 7RH, United Kingdom\\
Cavendish Laboratory, University of Cambridge,
Cambridge, CB3 0HE, United Kingdom}

\begin{abstract}

The internal electroluminescent quantum efficiency of organic
light emitting diodes is largely determined by the yield of
singlet excitons formed by the recombination of the injected
electrons and holes. Many recent experiments indicate that in
conjugated polymer devices this yield exceeds the statistical
limit of 25\% expected when the recombination is spin-independent.
This paper presents a possible explanation for these results. We propose a theory of electron-hole recombination via inter-molecular inter-conversion from inter-molecular weakly bound polaron pairs (or charge-transfer excitons) to intra-molecular excitons. This theory is applicable to parallel polymer chains. A crucial aspect of the theory is that both the intra-molecular and  inter-molecular excitons are effective-particles, which are described by both a relative-particle wavefunction and a center-of-mass wavefunction. This implies two electronic selection rules. (1) The parity of the relative-particle wavefunction implies that inter-conversion occurs from the even parity inter-molecular charge-transfer excitons to the strongly bound intra-molecular excitons. (2) The orthonormality of the center-of-mass wavefunctions ensures that inter-conversion occurs from the charge-transfer excitons to the lowest branch of the strongly bound exciton families, and not to higher lying members of these families. The inter-conversion is then predominately a multi-phonon process,  determined by the Franck-Condon factors. These factors are exponentially smaller for the triplet manifold than the singlet manifold because of the large exchange energy.
As a consequence, the inter-conversion rate in the triplet manifold is significantly smaller than that of the singlet manifold, and indeed it is comparable to the inter-system crossing rate. Thus, it is possible for the singlet exciton yield in conjugated polymers to be considerably enhanced over the spin-independent recombination value of $25\%$

\end{abstract}


\maketitle

\section{Introduction}

The internal electroluminescent quantum efficiency of organic
light emitting diodes is largely determined by the yield of
singlet excitons formed by the recombination of the injected
electrons and holes. Singlet exciton yields in light emitting
polymers exceeding the spin-independent recombination value of
25\% have now been reported by a large number of groups
\cite{1}-\cite{7}, although its value remains controversial. Lin
\textit{et al.}\ \cite{8} claim that the singlet yield only
exceeds the statistical limit
 in large electric fields,\cite{foot1} while Segal \textit{et al.}\ \cite{9} report
 a singlet exciton yield of only 20\%. A photo-luminescence detected
 magnetic resonance investigation\cite{17a} suggests that inter-chain (or bimolecular) recombination is spin-dependent.

 Many theoretical attempts have  been made to explain the enhanced singlet exciton yield.
 Bittner \textit{et al.}\ \cite{10}-\cite{12} assume that intra-chain electron-hole
 recombination occurs via vibrational relaxation through the band of exciton states between the
 particle-hole continuum and lowest bound excitons.
 Since vibrational relaxation
 is faster in the singlet channel than the triplet channel, because the lowest singlet exciton
 lies higher in energy
 than the lowest triplet exciton,
 a faster formation rate for the singlet than the triplet exciton is predicted. Hong and Meng
 \cite{13}
 argue that a multi-phonon process in the triplet channel also leads to faster intra-molecular singlet
 exciton formation.

 The different rates for singlet and triplet exciton formation predicted in the literature for inter-chain recombination \cite{14}-\cite{16} arise
 largely from the assumption that an inter-chain density-dependent electron transfer term is an important
 factor in the recombination mechanism. This term couples states of the same ionicity. 
 Since the inter-chain charge transfer states are predominately ionic, while the intra-chain
 triplet exciton has more covalent character than the intra-chain singlet exciton, the rate
  for the singlet exciton formation is correspondingly greater.

 Most of the recent experimental
 and theoretical work is reviewed in refs\cite{17,18}.

In this paper we develop a  model of inter-chain electron-hole
recombination between pairs of parallel polymers that involves intermediate, loosely bound
(`charge-transfer') states that lie energetically between the
electron-hole continuum and the final, strongly bound exciton
states. We argue that as a consequence of electronic selection rules, inter-molecular inter-conversion occurs from the
charge transfer to the lowest energy exciton states. This process is then limited by multi-phonon
emission, which decreases approximately exponentially with the
energy gap between the pair of states. Since the lowest singlet
and triplet exciton energies are split by a large exchange energy
(of ca.\ 0.7 eV\cite{kohler}), while the charge-transfer states are
quasi-degenerate, the triplet exciton formation rate is
considerably smaller than the singlet exciton rate. 

Multi-phonon emission has already been discussed as a possible
factor in determining the overall singlet exciton yield in intra-molecular processes\cite{19,13}. However, our model differs from these works by
emphasizing the important role of the intermediate inter-chain
charge-transfer states.

In the next section we introduce the relevant rate equations and derive an expression for the singlet exciton
yield as a function of the characteristic relaxation times. In
the following  section the microscopic model of  inter-molecular
inter-conversion is described and the  inter-conversion
rates are calculated. 

\section{Basic model and the rate equations}

\begin{figure}[tb]
\begin{center}
\includegraphics[scale=0.60]{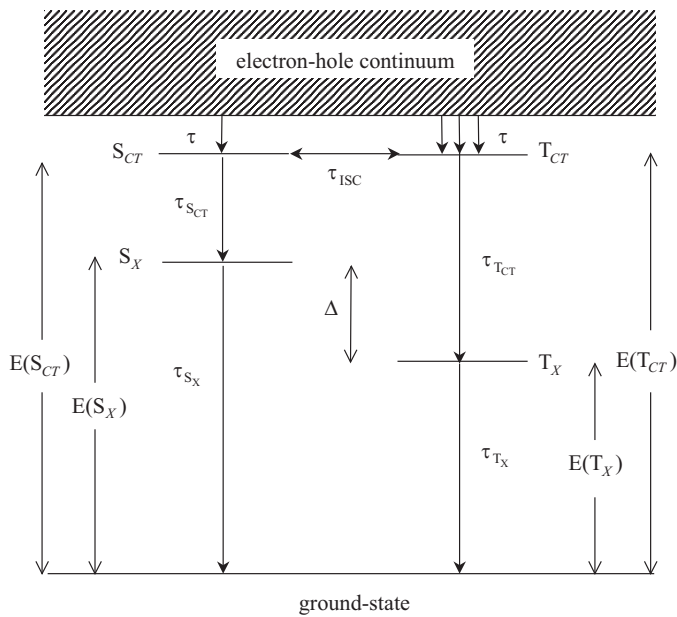}
\end{center}
\caption{The energy level diagram of the quasi-degenerate singlet and triplet
 charge-transfer excitons (denoted by $S_{CT}$ and $T_{CT}$,
respectively) and the lowest singlet and triplet excitons (denoted
by $S_{X}$ and $T_{X}$, respectively). $S_{CT}$ and $T_{CT}$ may be either intra-molecular odd parity excitons or inter-molecular even parity excitons. (In each case these
correspond to the lowest pseudo-momentum members of each exciton
family, as described in Section III.) Also shown are the respective
lifetimes (or inverse rates) for the inter-conversions within the
same spin manifolds and inter-system crossing (ISC) between the spin
manifolds. $\Delta$ is the exchange energy between $S_X$ and
$T_X$. } \label{Fi:10.1}
\end{figure}

Fig.\ 1 shows the energy level diagram for this
model. The electrons and holes are injected into the polymer
device with random spin orientations. Under the influence of the
electric field the electrons and holes migrate through the device,
rapidly being captured (in less than $10^{-12}$ s) to form the
weakly bound charge-transfer singlet and triplet excitons,
$S_{CT}$ and $T_{CT}$, respectively. We assume that no spin mixing
occurs during this process, and thus the ratio of $S_{CT}$ to
$T_{CT}$ is 1:3.\cite{foot3} If the inter-system-crossing (ISC) between
$T_{CT}$ and $S_{CT}$ (with a rate $1/\tau_{ISC}$) competes with the 
inter-conversion from $T_{CT}$ to the triplet exciton, $T_X$,
(with a rate $1/\tau_{T_{CT}}$) and $1/\tau_{T_{CT}}$ is smaller than 
the inter-conversion rate ($1/\tau_{S_{CT}}$) from $S_{CT}$ to the
singlet exciton, $S_X$, then the singlet yield is enhanced.

The charge-transfer states might be either intra-molecular
loosely bound excitons or weakly bound positive and negative
polarons on neighboring chains. Intra-molecular charge-transfer
states are Mott-Wannier excitons whose relative electron-hole
wavefunctions are odd under electron-hole exchange\cite{20}. A
crucial characteristic of these states is that, because of their
odd electron-hole parity, the probability of finding the electron
and hole on the same molecular repeat-unit is zero. Thus, they
experience very small exchange interactions and therefore the
singlet and triplet states are quasi-degenerate\cite{20},\cite{foot2}. 
Similarly, the inter-molecular weakly bound positive and
negative polarons - although now possessing even electron-hole
parity - also experience weak exchange interactions as necessarily
the electron and hole are on different repeat units, and thus
singlet and triplet states are also quasi-degenerate.
Furthermore,
since the charge-transfer excitons are also weakly bound with relatively large
electron-hole separations, there exist efficient spin-flipping
mechanisms, such as spin-orbit coupling, or exciton dissociation
via the electric field or by scattering from free carriers and
defects. 
In this paper we  focuss on inter-conversion to the intra-molecular  excitons from the inter-chain charge-transfer excitons.

The strongly bound excitons, $S_X$ and $T_X$, are intra-molecular
states. The inter-conversion process from $S_{CT}$ to $S_X$ and
from $T_{CT}$ to $T_X$ depends on the nature of $S_{CT}$ and
$T_{CT}$. The mechanism for bi-molecular inter-conversion is described more fully in the next
section.  In this section we describe the kinetics by
classical rate equations. The use of classical rate equations is
justified if rapid inter-conversion follows the ISC between
$T_{CT}$ and $S_{CT}$, as then there will be no coherence or
recurrence between $T_{CT}$ and $S_{CT}$\cite{24}. We also note
that since inter-conversion is followed by rapid
vibrational-relaxation (in a time of $\sim 10^{-13}$ s) these
processes are irreversible.

We first consider the case where ISC occurs directly via the spin-orbit coupling operator. This operator 
converts the $S_z = \pm 1$ triplets into the singlet\cite{13}, and \textit{vice versa}.
Let $N_{S_X}$, $N_{S_{CT}}$, $N_{T_X}^{\pm}$ and $N_{T_{CT}}^{\pm}$ denote the
number of $S_X$, $S_{CT}$, and the $S_z=\pm 1$ $T_X$ and $T_{CT}$ excitons,
respectively. $N/\tau$ is the number electron-hole pairs
created per second. Then the rate equations are:
\begin{equation}\label{}
    \frac{d N_{S_{CT}}} {dt} = \frac{N}{4\tau} +
    \frac{N_{T_{CT}}^{\pm}} {\tau_{ISC}} - N_{S_{CT}} \left( \frac{1}{\tau_{ISC}}
    + \frac{1}{\tau_{S_{CT}}} \right),
\end{equation}
\begin{equation}\label{Eq:2}
    \frac{d N_{T_{CT}}^{\pm}} {dt} = \frac{ N}{2\tau} +
    \frac{N_{S_{CT}}} {\tau_{ISC}} - N_{T_{CT}}^{\pm} \left( \frac{1}{\tau_{ISC}}
    + \frac{1}{\tau_{T_{CT}}} \right),
\end{equation}
\begin{equation}\label{}
    \frac{d N_{S_X}} {dt} = \frac{N_{S_{CT}}}{\tau_{S_{CT}}} -
    \frac{N_{S_X}} {\tau_{S_X}} ,
\end{equation}
and
\begin{equation}\label{}
    \frac{d N_{T_X}^{\pm}} {dt} = \frac{N_{T_{CT}}^{\pm}}{\tau_{T_{CT}}} -
    \frac{N_{T_X}^{\pm}} {\tau_{T_X}}.
\end{equation}
Notice that the $S_z = 0$ component of the
$T_{CT}$ exciton is \textit{converted directly} to the $S_z = 0$ component of the $T_X$ exciton, and cannot contribute to the singlet exciton yield.

When these equations are solved under the steady state conditions
that
\begin{equation}\label{}
 \frac{d N_{S_{CT}}} {dt}  =  \frac{d N_{T_{CT}}^{\pm}} {dt}  =  \frac{d N_{S_X}}
 {dt}=  \frac{d N_{T_X}^{\pm}} {dt}=0
\end{equation}
we obtain the singlet exciton yield, $\eta_S$, defined by,
\begin{equation}\label{}
    \eta_S=\frac{N_{S_X}/\tau_{S_X}}{N_{S_X}/\tau_{S_X} +
    N_{T_X}/\tau_{T_X}} \equiv \frac{N_{S_X}/\tau_{S_X}}{N/\tau},
\end{equation}
as
\begin{equation}\label{Eq:7}
    \eta_S=\frac{3+\gamma}{4(1+\beta+\gamma)},
\end{equation}
where  $\beta = \tau_{S_{CT}}/\tau_{T_{CT}}$ and $\gamma =
\tau_{ISC}/\tau_{T_{CT}}$. 

Alternatively, we might consider ISC via a spin-randomization process, whereby the charge-transfer excitons are scattered into charge-transfer triplets with a probability of $3/4$ and charge-transfer singlets with a probability of $1/4$. Then the rate equations are those of the Appendix of ref\cite{17} and the singlet exciton yield becomes\cite{17},
\begin{equation}\label{Eq:7b}
    \eta_S=\frac{1+\gamma}{1+3\beta + 4\gamma},
\end{equation}

\begin{figure}[tb]
\begin{center}
\includegraphics[scale=0.60]{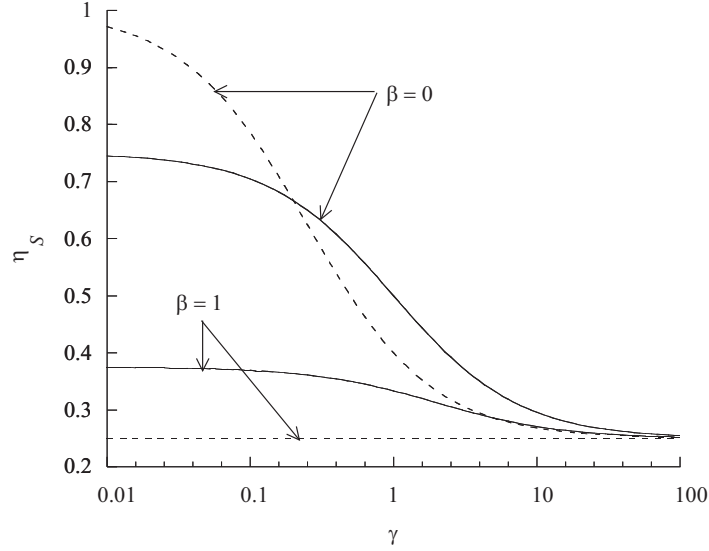}
\end{center}
\caption{The singlet exciton yield, $\eta_S$, versus $\gamma = \tau_{ISC}/\tau_{T_{CT}}$. $\beta = \tau_{S_{CT}}/\tau_{T_{CT}}$. Solid curves from Eq.\ (\ref{Eq:7}), dashed curves from Eq.\ (\ref{Eq:7b}).} \label{Fi:2}
\end{figure}

In practice, as we shall show,  $\tau_{T_{CT}} \gg \tau_{S_{CT}}$
so $\beta \approx 0$. We note that $\eta_S$ is a function only of
the relative life-times of $S_{CT}$ and $T_{CT}$, and the ISC
rate.  The singlet yield is plotted in Fig.\ \ref{Fi:2} as a function of $\gamma$. We now describe the calculation of the relative rates.

\section{Derivation of the inter-molecular inter-conversion rate}

The Born-Oppenheimer (B-O) Hamiltonian for a pair of coupled
polymer chains is,
\begin{equation}\label{}
    H = \sum_{\ell=1}^2 H_{intra}^{(\ell)} + H_{inter},
\end{equation}
where $H_{intra}^{(\ell)}$ is the intra-chain B-O Hamiltonian for
the $\ell$th chain and $H_{inter}$ is the inter-chain B-O
Hamiltonian. We split the inter-chain Hamiltonian into two
components: the inter-chain one-electron Hamiltonian,
$H_{inter}^1$, and the inter-chain two-electron Hamiltonian,
$H_{inter}^2$. $H_{inter}^2$ predominately describes the Coulomb
interactions between the $\pi$-electrons on neighboring chains.
$H_{inter}^1$ describes electron transfer between chains. For parallel chains with nearest neighbor electron transfer this is,
\begin{equation}\label{}
    H_{inter}^1 = - t_{inter} \sum_{i \sigma} \left(
    c_{i\sigma}^{(1)\dagger} c_{i\sigma}^{(2)} +
    c_{i\sigma}^{(2)\dagger}c_{i\sigma}^{(1)}\right),
\end{equation}
where $c_{i\sigma}^{(\ell)\dagger}$  ($c_{i\sigma}^{(\ell)}$)
creates (destroys) a $\pi$-electron on site $i$ of chain $\ell$
and $t_{inter}$ is the inter-chain hybridization integral. If the
chains are weakly coupled we may regard $H_{inter}^1$ as a
perturbation on the approximate Hamiltonian,
\begin{equation}\label{}
    {\tilde H} = \sum_{\ell=1}^2 H_{intra}^{(\ell)} + H_{inter}^2.
\end{equation}

Within the Born-Oppenheimer  approximation the electronic and nuclear degrees of
freedom are described by the Born-Oppenheimer states. A
Born-Oppenheimer state, $|A \rangle$, is a direct product of an
electronic state,  $|a; \{Q\} \rangle$, and a nuclear state
associated with that electronic state, $|\nu_a\rangle$:
\begin{equation}\label{}
    |A \rangle = |a; \{Q\} \rangle |\nu_a \rangle.
\end{equation}
The $\{Q\}$ label indicates that the electronic state is
parametrized by the nuclear coordinates.

The stationary electronic states are the eigenstates of the
approximate Hamiltonian, ${\tilde H}$. Thus, the perturbation,
$H_{inter}^1$ mixes these electronic states. In particular, it
causes an inter-conversion from the inter-chain excitons (or
weakly bound polaron pairs) to the intra-chain excitons by transferring charge from one chain to another.

We take the initial electronic state to be a positive polaron on
chain $1$ and a negative polaron on chain $2$,
\begin{equation}\label{}
   |i\rangle= |P^+,P^-;Q_1,Q_2\rangle.
\end{equation}
The inter-chain Coulomb interaction between the chains creates a weakly bound
charge-transfer exciton, to be described below. The labels $Q_1$
and $Q_2$ indicate the independent normal coordinates of chains 1
and 2, respectively.

We consider the situation where the negative polaron is
transferred from chain 2 to chain 1 by $H_{inter}^1$. Thus, the final state is an
intra-molecular exciton on chain 1 (denoted by $|a\rangle$),
leaving chain 2 in its ground electronic state,
\begin{equation}\label{}
   |f\rangle= |a;Q_1\rangle^{(1)}|1A_g;Q_2\rangle^{(2)}.
\end{equation}

Before proceeding it will be useful to review the theory of excitons
in conjugated polymers. In the weak-coupling limit (namely, the limit that the
Coulomb interactions are less than or equal to the band width) the
intra-molecular excited states of semi-conducting conjugated
polymers are Mott-Wannier excitons described by\cite{20},
 \begin{equation}\label{Eq:9}
    |a;Q \rangle =  \sum_{r,R}
    \psi_n(r;Q)\Psi_j(R) |r,R \rangle.
\end{equation}
$|r,R \rangle$ is an electron-hole basis state constructed by
promoting an electron from the filled valence band Wannier orbital
at $R - r/2$ to the empty conduction band Wannier orbital at $R +
r/2$,
\begin{equation}\label{Eq:15}
  |r,R \rangle = \frac{1}{\sqrt{2}} \left( c_{R+r/2, \uparrow}^{c (\ell)\dagger} c_{R-r/2,
  \uparrow}^{v(\ell)}
  \pm c_{R+r/2, \downarrow}^{c (\ell)\dagger} c_{R-r/2, \downarrow}^{v(\ell)} \right)|1^1A_g\rangle.
\end{equation}
$c_{m\sigma}^{v(\ell)\dagger}$ and
$c_{m\sigma}^{c(\ell)\dagger}$ are the valence and conduction
Wannier orbital operators, respectively, approximately defined by,
\begin{equation}\label{}
    c_{m\sigma}^{v(\ell)\dagger} = \frac{1}{\sqrt{2}} \left(c_{2m-1\sigma}^{(\ell)\dagger}
    + c_{2jm\sigma}^{(\ell)\dagger}\right)
\end{equation}
and
\begin{equation}\label{}
    c_{m\sigma}^{c(\ell)\dagger} = \frac{1}{\sqrt{2}} \left(c_{2m-1\sigma}^{(\ell)\dagger}
    - c_{2m\sigma}^{(\ell)\dagger}\right),
\end{equation}
where $m$ is the unit cell index. The $\pm$ symbol in Eq.\ (\ref{Eq:15})
refers to singlet ($+$) or triplet ($-$) excitons.

$R$ is the center-of-mass coordinate and $r$ is the relative
coordinate of the effective particle. $\psi_n(r;Q)$ is a
hydrogen-like electron-hole wavefunction labelled by the principle
quantum number, $n$, which describes the effective-particle. This
has the property that under electron-hole reflection (namely, $r
\rightarrow -r$) $\psi_n(r;Q) = \psi_n(-r;Q)$ for odd $n$ and
$\psi_n(r;Q) = -\psi_n(-r;Q)$  for even $n$. 
\begin{equation}\label{Eq:com}
   \Psi_j(R) =  \sqrt{\frac{2}{N+1}}
\sin(\beta R)
\end{equation}
is the center-of-mass wavefunction, which describes
the motion of the effective-particle on a linear chain. $N$ is the
number of unit cells. For each principle quantum number, $n$,
there is a \emph{band} of excitons with different pseudo-momentum,
$\beta = \pi j/(N+1)d$, where $j$ satisfies $ 1 \leq j \leq N$ and
$d$ is the unit cell distance. Thus, every exciton state label,
$a$, corresponds to two independent quantum numbers: $n$ and $j$.
As described in\cite{20}, $n = 1$ corresponds to the $S_X$ and
$T_X$ families of intra-chain excitons, while $n = 2$ corresponds
to the $S_{CT}$ and $T_{CT}$ families of intra-chain excitons. The
lowest energy member of each family has the smallest
pseudo-momentum, namely, $j = 1$\cite{foot10}.

It is also convenient to describe the  inter-molecular weakly
bound polaron pairs as charge-transfer excitons described by,
 \begin{equation}\label{Eq:18}
    |P^+,P^-;Q_1,Q_2 \rangle =  \sum_{r,R}
    {\tilde \psi}_n(r;Q_1,Q_2)\Psi_j(R) |r,R;2 \rangle,
\end{equation}
where  ${\tilde \psi}_n$ represents the inter-chain
effective-particle  wavefunction. $n=1$ (i.e.\ even electron-hole parity) for the lowest energy inter-chain excitons. $|r,R;2 \rangle$ is an
electron-hole basis state constructed by promoting an electron
from the filled valence band Wannier orbital at $R - r/2$ on chain
1 to the empty conduction band Wannier orbital at $R + r/2$ on
chain 2,
\begin{equation}
  |r,R;2 \rangle = \frac{1}{\sqrt{2}} \left( c_{R+r/2, \uparrow}^{c (2) \dagger} c_{R-r/2, \uparrow}^{v (1)}
  \pm c_{R+r/2, \downarrow}^{c (2) \dagger} c_{R-r/2, \downarrow}^{v(1)}
  \right)|1^1A_g\rangle^{(1)}|1^1A_g\rangle^{(2)}.
\end{equation}

With this background to the theory of excitons we now proceed to derive the
transfer rate. The iso-energetic inter-conversion rate from the
initial to the final states is determined by the Fermi Golden Rule
expression,
\begin{equation}\label{Eq:20a}
    k_{I \rightarrow F} = \frac{2\pi}{\hbar} \langle F |H_{inter}^1 |
    I \rangle^2 \delta(E_F-E_I),
\end{equation}
where the initial and final B-O states are,
\begin{equation}\label{19}
    |I\rangle = |P^+,P^-;Q_1,Q_2\rangle|\nu_{P^+}\rangle^{(1)}|\nu_{P^-}\rangle^{(2)}
\end{equation}
and
\begin{equation}\label{20}
    |F\rangle = |a;Q_1\rangle^{(1)}|1A_g;Q_2\rangle^{(2)}
    |\nu_{a}\rangle^{(1)}|\nu_{1^1A_g}\rangle^{(2)},
\end{equation}
respectively.

\subsection{Electronic matrix elements}

The corresponding electronic matrix element is,
\begin{eqnarray}\label{}
   \langle f |H_{inter}^1 |
    i \rangle = &&
 ^{(2)}\langle 1^1A_g;Q_2|^{(1)}\langle a;Q_1|H_{inter}^1
 |P^+,P^-;Q_1,Q_2\rangle.
\end{eqnarray}
Using Eq.\ (\ref{Eq:9}) and Eq.\ (\ref{Eq:18}) this is,
\begin{eqnarray}\label{}
   \langle f |H_{inter}^1 |
    i \rangle
 = &&
\left({\frac{2}{N+1}}\right)
 \sum_{r',R'} \psi_{n'}(r';Q_1)\sin(\beta' R') \sum_{r,R}
    {\tilde \psi}_n(r;Q_1,Q_2)\sin(\beta R) \times
    \nonumber
    \\
    &&
    ^{(2)}\langle 1^1A_g;Q_2| ^{(1)}\langle r', R' |H_{inter}^1|r,R;2
    \rangle.
\end{eqnarray}
This  matrix element is evaluated by expressing $H_{inter}^1$ in
terms of the valence
 and conduction
 Wannier orbital operators. Retaining terms that keep within the exciton sub-space we have,
\begin{equation}\label{}
    H_{inter}^1 = - t_{inter} \sum_{m \sigma} \left(
    c_{m\sigma}^{v(1)\dagger} c_{m\sigma}^{v(2)} +
    c_{m\sigma}^{c(1)\dagger}c_{m\sigma}^{c(2)}\right) +
    \textrm{H.C.}
\end{equation}
Then,
\begin{eqnarray}\label{}
    &&  \langle f|H_{inter}^1|i\rangle  =
    \nonumber
    \\
    &&
    -t_{inter} \left({\frac{2}{N+1}}\right)\sum_{r',r}\psi_{n'}(r';Q_1)\tilde{\psi}_n(r;Q_1,Q_2)
    \sum_{R',R}
    \sin(\beta' R') \sin(\beta R) ^{(1)}\langle
    r',R'|r,R\rangle^{(1)}.
\end{eqnarray}
By  exploiting the orthonormality of the basis functions,
\begin{equation}\label{}
\langle r',R'|r,R\rangle = \delta_{r' r}\delta_{R'R},
\end{equation}
as well as the $\sin(\beta R)$ functions,
\begin{equation}\label{Eq:28}
\left({\frac{2}{N+1}}\right) \sum_{R}
    \sin(\beta' R) \sin(\beta R) = \delta_{\beta' \beta},
\end{equation}
we have the final result for the electronic matrix element,
\begin{eqnarray}\label{Eq:29}
    \langle f|H_{inter}^1|i\rangle =
 -t_{inter} \sum_{r}\psi_{n'}(r;Q_1)\tilde{\psi}_n(r;Q_1,Q_2).
\end{eqnarray}

Eq.\ (\ref{Eq:28}), Eq.\ (\ref{Eq:com}) and Eq.\ (\ref{Eq:29}) demonstrate the very significant result that interconverion via
$H_{inter}^1$ is subject to two electronic selection rules.
\begin{enumerate}
\item{Inter-conversion occurs between excitons with the same  centre-of-mass pseudo-momentum, $\beta_j$.}
\item{Inter-conversion occurs between excitons with the same  electron-hole parity. Thus, $|n'-n| =\textrm{ even}$.}
\end{enumerate}
Since the lowest energy inter-chain
excitons have even electron-hole parity this implies
that $H_{inter}^1$ connects them to $S_X$ and $T_X$, and not to
the intra-molecular $S_{CT}$ and $T_{CT}$.\cite{foot6} Moreover,
since the inter-chain exciton will have relaxed to its lowest momentum
state, $H_{inter}^1$ converts it to the intra-chain exciton in its
lowest momentum state, and \emph{not to higher lying momentum
states}.

\subsection{Vibrational overlaps}

We now discuss the contribution of the vibrational wavefunctions to the total matrix element.
Inter-molecular inter-conversion is an iso-energetic process
which  occurs from the lowest pseudo-momentum state of the
charge-transfer manifold and the lowest vibrational levels of this
state to the lowest pseudo-momentum state of the intra-molecular
excitons  at the same energy as the initial level. Thus, the
vibrational levels in Eq.\ (\ref{19}) are $\nu_{P^+} = 0$ and
$\nu_{P^-} = 0$. However, the vibrational levels in Eq.\
(\ref{20}) are determined by the conservation of energy.

\begin{figure}[tb]
\begin{center}
\includegraphics[scale=0.60]{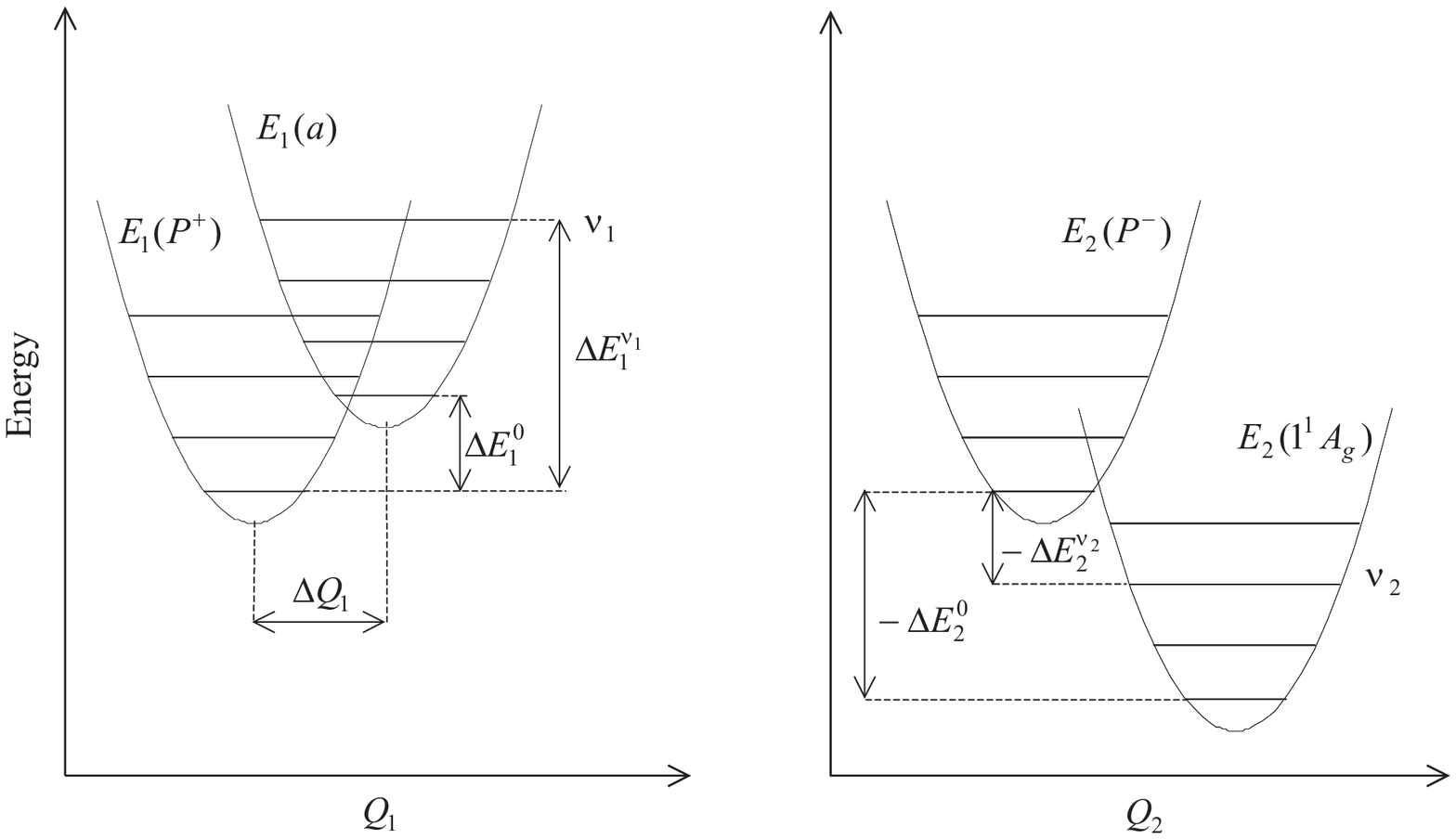}
\end{center}
\caption{Bimolecular electron transfer from chain 2 to chain 1.
The initial state is a positive polaron on chain 1 and a negative
polaron on chain 2, each in their lowest vibrational level. The
electron transfer creates an exciton state in chain 1, with chain
2 in its ground state. The energy differences between the final
and initial states are $\Delta E^{\nu_{\ell}}_{\ell}$ for the
$\ell$th chain. If $\Delta Q_1 = 0$ then $\Delta E^{\nu_{1}}_{1} =
\Delta E^{0}_{1}$ and thus $\Delta E^{\nu_{2}}_{2} = -\Delta
E^{0}_{1}$.} \label{Fi:10.3}
\end{figure}

Using Eq.\ (\ref{Eq:20a}) the rate is thus,
\begin{eqnarray}\label{Eq:30}
    k_{I \rightarrow F} && = \frac{2 \pi}{\hbar}
    |\langle f|H_{inter}^1|i\rangle|^2\sum_{\nu_1 \nu_2}
    F_{0\nu_1}^{(1)}F_{0\nu_2}^{(2)}\delta( E_F - E_I)
    \nonumber
    \\
&& = \frac{2 \pi}{\hbar}
    |\langle f|H_{inter}^1|i\rangle|^2\sum_{\nu_1 \nu_2}
    F_{0\nu_1}^{(1)}F_{0\nu_2}^{(2)}\delta(\Delta E_1^{\nu_1} + \Delta
    E_2^{\nu_2})
\end{eqnarray}
where
\begin{equation}\label{}
F_{0\nu_1}^{(1)} = |^{(1)}\langle 0_{P^+}|\nu_{a}\rangle^{(1)}|^2
\end{equation}
and
\begin{equation}\label{}
F_{0\nu_2}^{(2)} = |^{(2)}\langle
0_{P^-}|\nu_{1^1A_g}\rangle^{(2)}|^2
\end{equation}
are the Franck-Condon factors associated with the vibrational
wavefunction overlaps of chains 1 and 2, respectively. Likewise,
\begin{equation}\label{}
\Delta E_1^{\nu_1} = E_1(a; \nu_a) - E_1(P^+;0_{P^+})
\end{equation}
and
\begin{equation}\label{}
\Delta E_2^{\nu_1} = E_2(1^1A_g; \nu_{1^1A_g}) - E_2(P^-;0_{P^-})
\end{equation}
are the changes in energy of chains 1 and 2, respectively. These
changes in energy are illustrated in Fig.\ \ref{Fi:10.3}.

Using the identity,
\begin{equation}\label{}
\delta(x+y) = \int \delta(x-z) \delta(y+z)dz
\end{equation}
we can re-write Eq.\ (\ref{Eq:30}) as,
\begin{equation}\label{Eq:36}
    k_{I \rightarrow F} = \frac{2 \pi}{\hbar}
    |\langle f|H_{inter}^1|i\rangle|^2 \int \sum_{\nu_1 \nu_2}
    F_{0\nu_1}^{(1)}F_{0\nu_2}^{(2)}\delta(\Delta E_1^{\nu_1} -E)\delta(\Delta
    E_2^{\nu_2} +E)dE.
\end{equation}
Defining the spectral functions for the donor (chain 2) and
acceptor (chain 1) as,
\begin{equation}\label{}
    D(E)  = \sum_{\nu_2}
   F_{0\nu_2}^{(2)}\delta(\Delta
    E_2^{\nu_2} +E)
\end{equation}
and
\begin{equation}\label{}
    A(E) = \sum_{\nu_1}
    F_{0\nu_1}^{(1)}\delta(\Delta E_1^{\nu_1} -E),
\end{equation}
respectively, we have the familiar rate expression for
bi-molecular electron transfer,
\begin{equation}\label{}
 k_{I \rightarrow F} = \frac{2 \pi}{\hbar}
    |\langle f|H_{inter}^1|i\rangle|^2 \int D(E)A(E)dE.
\end{equation}

A useful simplification to this expression arises by noting that
the geometric distortions of the polarons and exciton polarons
(namely the $1^1B_u$ or $1^3B_u$ states) from the ground state
structure are very similar.\cite{32} Thus, the Huang-Rhys parameter (proportional to $\Delta Q_1^2$, as defined in Fig.\ 3)
for the $1^1B_u$ and $1^3B_u$ states relative to the positive
polaron is negligible. Therefore,
\begin{equation}\label{}
    F_{0\nu_1} \sim \delta_{0\nu_1},
\end{equation}
and thus the change of energy of chain 1 is,
\begin{equation}\label{Eq:41}
    \Delta E_1^{\nu_1} \equiv   \Delta E_1^{0},
\end{equation}
where $\Delta E_1^{0}$ is the $0-0$ energy difference on chain 1
between the final exciton state and the positive polaron. This is
illustrated in Fig.\ 3. By the conservation of energy we
therefore have,
\begin{equation}\label{}
    \Delta E_2^{\nu_2}  = - \Delta E_1^{0}.
\end{equation}

The vibrational level, $\nu_2$, of the final $1^1A_g$ state of
chain 2 to which inter-conversion from the negative polaron
initially occurs is given by,
\begin{eqnarray}\label{Eq:43}
    \nu_2 && = (\Delta E_2^{0} -  \Delta E_2^{\nu_2})/\hbar\omega = (\Delta E_2^{0} + \Delta E_1^{0})/\hbar\omega
    \nonumber
    \\
    && = ( E_2(P^-;0_{P^-}) - E_2(1^1A_g;0_{1^1A_g}) + E_1(P^+;0_{P^+})- E_1(a;0_a) )/\hbar \omega
    \nonumber
    \\
  && =  (\Delta E_{CT} - \Delta E_a)/\hbar \omega,
\end{eqnarray}
where
\begin{equation}\label{}
\Delta E_{CT} = (E_1(P^+;0_{P^+})-E_1(1^1A_g;0_{1^1A_g})) +
(E_2(P^-;0_{P^-})- E_2(1^1A_g;0_{1^1A_g}))
\end{equation}
 and
\begin{equation}\label{}
\Delta E_{a} = E_1(a;0_a) - E_1(1^1A_g;0_{1^1A_g})
\end{equation}
are the $0-0$ transition energies of the  charge-transfer exciton
and the state $|a\rangle$, respectively.

The condition expressed in Eq.\ (\ref{Eq:41}) implies that the energy
integral in Eq.\ (\ref{Eq:36}) is restricted to the value of $E = \Delta
E_1^0$, and thus the rate becomes,
\begin{eqnarray}\label{}
    k_{I \rightarrow F} && = \frac{2 \pi}{\hbar}
    |\langle f|H_{inter}^1|i\rangle|^2 \sum_{\nu_2}
  F_{0\nu_2}^{(2)}\delta(\Delta
    E_2^{\nu_2} + \Delta E_1^0)
    \nonumber
    \\
&& = \frac{2 \pi}{\hbar}
    |\langle f|H_{inter}^1|i\rangle|^2
  F_{0\nu_2}^{(2)}\rho_f(E).
\end{eqnarray}
$\rho_f(E)$ is the final density of states on chain 2, defined by,
\begin{equation}\label{}
\rho_f(E) = \sum_{\nu_2}\delta(\Delta
    E_2^{\nu_2} + \Delta E_1^0),
\end{equation}
which is usually taken to be the inverse of the vibrational energy
spacing. Inserting the expression for the Franck Condon factor,
\begin{equation}\label{}
      F_{0\nu_2}^{(2)} = |^{(2)}\langle
0_{P^-}|\nu_{1^1A_g}\rangle^{(2)}|^2 =
\frac{\exp(-\textsf{S}_p)
\textsf{S}_p^{\nu_2}}{\nu_2!}
\end{equation}
we have the final result that,
\begin{equation}\label{}
    k_{I \rightarrow F}  = \frac{2 \pi}{\hbar}
    |\langle f|H_{inter}^1|i\rangle|^2 \rho_f(E)
\frac{\exp(-\textsf{S}_p) \textsf{S}_p^{\nu_2}}{\nu_2!}.
\end{equation}
This equation, along with the definition of $\langle
f|H_{inter}^1|i\rangle$ in Eq.\ (\ref{Eq:29}), is our final expression for
the inter-conversion rate. $\textsf{S}_p$ is the Huang-Rhys factor
for the polaron relative to the ground state, defined as
$E_d/\hbar\omega$, where $E_d$ is the reorganization (or
relaxation) energy of the polaron relative to the ground state.
After the iso-energetic transition there is vibrational relaxation
to the lowest vibrational level of the $1^1A_g$ state of chain 2
via the sequential emission of $\nu_2$ phonons. The number of
phonons emitted corresponds to the difference in energies between
the initial charge-transfer and final exciton states, given by Eq.\ (\ref{Eq:43}). This is a
multi-phonon process. In the next section we  estimate these rates.

\section{Estimate of the inter-conversion rates}

Since inter-conversion from the inter-molecular to the intra-molecular charge-transfer excitons is forbidden by symmetry, we now only discuss inter-conversion to the lowest excitons, $S_X$ or $T_X$. (As remarked in footnote\cite{foot6}, inter-conversion to higher-lying exciton states is allowed, but if this happens recombination is an intra-molecular process via the intra-molecular charge-transfer excitons.) Thus, the state label $a$ is now either $S_X$ or $T_X$, and the number of phonons emitted, $\nu_2$, is either $\nu_S$ or $\nu_T$, as determined by Eq.\ (\ref{Eq:43}).

Within the Mott-Wannier basis the exciton wavefunction overlaps
are easy to calculate. Using  $t_{inter} = 0.1$ eV\cite{16}, the
interchain distance as $4$ $\AA$ and standard semi-empirical Coulomb
interactions gives
\begin{equation}\label{}
    \sum_r
    \psi_{S_X}(r)\tilde{\psi}_{S_{CT}}(r) \approx 1.0
\end{equation}
and
\begin{equation}\label{}
    \sum_r
    \psi_{T_X}(r)\tilde{\psi}_{T_{CT}}(r) \approx 0.9.
\end{equation}
The polaron Huang-Rhys parameter, $\textsf{S}_p$, is not accurately known for light emitting polymers. However, we expect it to be similar to the $S_X$ exciton Huang-Rhys parameter. The relaxation energy of the $S_X$ exciton has been experimentally determined as $0.07$ eV in PPV\cite{liess}, with a similar value calculated for `ladder' PPP in ref\cite{moore}. From the figures in ref\cite{hertel}, we estimate the relaxation energy to be $0.12$ eV in ladder PPP (where the phenyl rings are planar) and $0.25$ eV in PPP (where the phenyl rings are free to rotate).  Thus, taking the relaxation energy as $0.1$ eV and $\hbar \omega = 0.2$ eV implies that $\textsf{S}_p \sim 0.5$.

Now, using $\hbar \omega = 0.2$ eV ($\equiv \rho_f^{-1}$),  $\textsf{S}_p \sim 0.5$ and assuming that  
the energy difference between the strongly bound singlet exciton ($S_X$) and the
\textit{intra-molecular} charge-transfer excitons of $\sim 0.8$ eV is approximately the energy difference between the singlet exciton and the
\textit{inter-molecular} charge-transfer excitons, we can estimate the inter-conversion
rate for the singlet exciton. This is $k_{S_{CT} \rightarrow S_X}
\sim 7.5 \times 10^{11}$ s$^{-1}$ (or $\tau_{S_{CT}} \sim 1$ ps). Similarly, using an exchange gap of $\sim
0.7$ eV gives $k_{T_{CT} \rightarrow T_X} \sim 1 \times 10^{8}$ s$^{-1}$ (or $\tau_{T_{CT}} \sim 10$ ns). Thus, the triplet inter-conversion rate is much slower than the singlet inter-conversion rate. 

The ISC rate is also not accurately known, with quoted values ranging from $0.3$ ns\cite{19}, $4$ ns\cite{frolov} and $10$ ns\cite{barford2004}. Nevertheless, despite this uncertainty, 
we see that the estimated triplet inter-conversion rate is  comparable to or slower than the ISC rate, which from Eq.\ (\ref{Eq:7})  implies a large singlet exciton yield.

Generally, the ratio of the rates is,
\begin{equation}\label{Eq:53}
    \frac{k_{S_{CT} \rightarrow S_X}}{k_{T_{CT} \rightarrow T_X}}=
   \frac{
    \left|\sum_r
    \psi_{S_X}(r)\tilde{\psi}_{S_{CT}}(r)
    \right|^2
\exp(-\textsf{S}_p) \frac{\textsf{S}_p^{\nu_S}}{\nu_S!}} {
\left|\sum_r
    \psi_{T_X}(r)\tilde{\psi}_{T_{CT}}(r)
    \right|^2
\exp(-\textsf{S}_p) \frac{\textsf{S}_p^{\nu_T }}{\nu_T!}}.
\end{equation}
Thus,
\begin{equation}\label{}
   \frac{k_{S_{CT} \rightarrow S_X}}{k_{T_{CT} \rightarrow T_X}}
  = 1.2 \textsf{S}_p^{-(\nu_T-\nu_S)}\frac{\nu_T !}{\nu_S!}.
\end{equation}
This ratio increases as $\textsf{S}_p$ decreases, because then multi-phonon emission becomes more difficult. The ratio also increases as the exchange energy, $\hbar\omega(\nu_T - \nu_S)$, increases for any $\nu_S$ or $\nu_T$ if $\textsf{S}_p < 1$.

\section{Discussion and Conclusions}

We propose a theory of electron-hole recombination via inter-molecular inter-conversion from inter-molecular weakly bound polaron pairs (or charge-transfer excitons) to  intra-molecular excitons. This theory is applicable to parallel polymer chains. A crucial aspect of the theory is that both the intra-molecular and  inter-molecular excitons are effective-particles, which are described by both a relative-particle wavefunction and a center-of-mass wavefunction. This implies two electronic selection rules.
\begin{enumerate}
\item{
The parity of the relative-particle wavefunction implies that inter-conversion occurs from the even parity inter-molecular charge-transfer excitons to the strongly bound intra-molecular excitons and not to the intra-molecular charge-transfer excitons (namely, the first odd parity exciton). (However, if the inter-chain charge transfer excitons lie higher in energy than the second family of even parity intra-chain exciton, recombination will be an intra-molecular process.)}
\item{The orthonormality of the center-of-mass wavefunctions ensures that inter-conversion occurs from the charge-transfer excitons to the lowest branch of the strongly bound exciton families, and not to higher lying members of these families.}
\end{enumerate}

These selection rules imply that inter-conversion is then predominately a multi-phonon process,  determined by the Franck-Condon factors. These factors are exponentially smaller for the triplet manifold than the singlet manifold because of the large exchange energy.

There is also a contribution to the rates from the overlap of the relative-particle wavefunctions, which again are smaller in the triplet manifold, because the triplet exciton has a smaller particle-hole separation and has more covalent character than its singlet counterpart\cite{16}. As a consequence, the inter-conversion rate in the triplet manifold is significantly smaller than that of the singlet manifold, and indeed it is comparable to the ISC rates. Thus, it is possible for the singlet exciton yield is expected to be considerably enhanced over the spin-independent value of $25\%$ in conjugated polymers.

Any successful theory must explain the observation that the singlet exciton yield is close to $25\%$ for molecules and increases with conjugation length\cite{3,5}. This theory qualitatively predicts this trend for two reasons. First, the effective-particle description of the exciton states is only formally exact for long chains. This description breaks down when the chain length (or more correctly, conjugation length) is comparable to the particle-hole separation. In this case separation of the center-of-mass motion and the relative-particle motion is no longer valid. Then the quantum numbers $n$ and $j$ (which describe the relative-particle wavefunction and center-of-mass wavefunction, respectively) are no longer independent quantum numbers. Inter-conversion is then expected to take place between all the states lying between the charge-transfer state and the lowest exciton state. However, as the chain length  increases inter-conversion to higher lying states is suppressed in favor of the lowest lying exciton. This prediction is confirmed by a recent quantum mechanical calcuations by Beljonne \textit{et al.} \cite{33}. The second reason that the singlet exciton yield is enhanced in polymers over molecules is that the Huang-Rhys parameters decrease as the conjugation length increases, and thus the relative rate (given by Eq. (\ref{Eq:53})) increases.

We note that the effective-particle description is still valid when there is self-trapping. In this case the center-of-mass wavefuctions are not the particle-in-the-box wavefunctions appropriate for a linear chains (Eq.\ (\ref{Eq:com})), but they are the ortho-normalized functions appropriate for the particular potential well trapping the effective particle. The key point is that because these are  ortho-normalized functions  inter-conversion occurs between a pair of states with the same center-of-mass quantum numbers, as described in this paper.

This theory has been formulated for an idealized case of sufficiently long, parallel polymer chains. The applicability of this theory for polymer light emitting displays needs verifying by performing calculations on oligomers of arbitrary length and arbitrary relative conformations.

Finally, we remark that this theory presents strategies for enhancing the singlet exciton yield. Ideally, the polymer chains should be well conjugated, closely packed and parallel. The last two conditions ensure that the inter-chain charge-transfer excitons lie energetically below high-lying even-parity families of intra-molecular excitons, and thus recombination is an inter-chain inter-conversion process and not an intra-molecular process via the intra-molecular charge-transfer excitons. Intra-molecular recombination is not desirable because although inter-conversion from the intra-chain charge-transfer excitons is slower in the triplet manifold than the singlet manifold, both rates are expected to be faster than the ISC rate. The relative inter-molecular interconversion rates are also increased when the electron-lattice coupling is reduced. This suggests that the singlet exciton yield is enhanced in rigid, well-conjugated polymers.

\begin{acknowledgements}
I thank  N. Greenham, A. K\"ohler, and C. Silva (Cambridge) for useful discussions.  I also gratefully acknowledge the financial support of the Leverhulme
Trust, and thank the Cavendish Laboratory and Clare Hall,
Cambridge for their hospitality.
\end{acknowledgements}


\begin{references}

\bibitem[]{email}
E.mail address: W.Barford@sheffield.ac.uk

\bibitem[]{} $^*$Permanent address.

\bibitem{1} Y. Cao, I. D. Parker, Y. Gang, C. Zhang, and A. J. Heeger, \textit{Nature} \textbf{397}, 414-417 (1999).

\bibitem{2} P. K. H. Ho, J.-S. Kim, J. H. Burroughes, H. Becker, F. Y. L. Sam,
 T. M. Brown, F. Cacialli, and F. H. Friend,
\textit{Nature} \textbf{404}, 481-484 (2000).

\bibitem{3} J. S. Wilson, A. S. Dhoot, A. J. A. B. Seeley, M. S. Khan,
A. K\"ohler, and R. H. Friend, \textit{Nature} 413, 828-831 (2001).

\bibitem{4} M. Wohlgennant, K. Tandon, S. Mazumdar, S. Ramasesha,
and Z. V. Vardeny. \textit{Nature} \textbf{409}, 494-497 (2001).

\bibitem{5} M. Wohlgennant, X. M. Jiang, Z. V. Vardeny,  and R. A. J.
Janssen, \textit{Phys. Rev.} \textbf{B88}, 197401 (2002).

\bibitem{6} M. Wohlgennant, C. Yang,  and Z. V. Vardeny,  \textit{Phys. Rev.}
\textbf{B66}, 241201-4 (2002).

\bibitem{7} A. S. Dhoot, D. S. Ginger, D. Beljonne, Z. Shuai,
and N. C. Greenham, \textit{Chem. Phys. Lett.} \textbf{360},
195-201 (2002).

\bibitem{8} L. C. Lin, H. F. Meng, J .T. Shy, S. F. Horng, L. S. Yu,
 C. H. Chen, H. H. Liaw, C. C. Huang, K. Y. Peng, and S. A. Chen,
\textit{Phys. Rev. Lett.} \textbf{90}, 36601 (2003).

\bibitem{foot1} These predictions have been questioned by R. {\"O}sterbacka,
\textit{Phys. Rev. Lett.}\textbf{ 91}, 219701, 2003,  A. S. Dhoot
and N. C. Greenham, \textit{ibid}, 219702, M. Schott,
\textit{Phys. Rev. Lett.}\textbf{ 92}, 59701, 2004, and W.
Barford, submitted.

\bibitem{9} M. Segal, M. A. Baldo, R. J. Holmes, S. R. Forrest,
and Z. G. Soos,  \textit{Phys. Rev.} \textbf{B68}, 075211 (2003).

\bibitem{17a} M. Wohlgennant, C. Yang, and Z. V. Vardeny, \textit{Phys.
Rev.} \textbf{B66}, 241210 (2002).

\bibitem{10} M. N. Kobrak and E. R. Bittner, \textit{Phys. Rev.} \textbf{B62},
11473-11486 (2000).

\bibitem{11} S. Karabunarliev  and E. R. Bittner, \textit{Phys. Rev. Lett.} \textbf{90},
057402 (2000).

\bibitem{12} S. Karabunarliev  and E. R. Bittner, \textit{J. Chem. Phys.}
\textbf{119}, 3988-3995 (2003).

\bibitem{13} T.-M. Hong and H.-F. Meng, \textit{Phys. Rev.} \textbf{B63}, 075206 (2001).

\bibitem{14} Z. Shuai, D. Beljonne, R. J. Silbey, and J. L. Bredas,
\textit{Phys. Rev. Lett.} \textbf{84}, 131-134 (2000).

\bibitem{15} A. Ye, Z. Shuai, and J. L.  Bredas, \textit{Phys. Rev.} \textbf{B65},
045208 (2002).

\bibitem{16} K. Tandon, S. Ramasesha, and S. Mazumdar, \textit{Phys. Rev.}
\textbf{B67}, 045109 (2003).

\bibitem{17} M. Wohlgennant and Z. V. Vardeny, \textit{J. Phys.: Condens.
Matter} \textbf{15}, R83-R107 (2003).

\bibitem{18} A. K\"ohler and J.  Wilson, \textit{Organic Electronics}, \textbf{4}, 179 (2003).

\bibitem{kohler} A. K\"ohler and D. Beljonne, \textit{Advanced Functional Materials}, \textbf{14}, 11, (2004).

\bibitem{19} A. L. Burin and M. A. Ratner, \textit{J. Chem. Phys.} \textbf{109},
6092-6102 (1998).

\bibitem{foot3} Furthermore, there is an efficient electric-field-induced
electron-hole mixing, so all the electron-hole pairs become
charge-transfer excitons.

\bibitem{20} W. Barford, R. J. Bursill, and R. W. Smith,  \textit{Phys. Rev.}
\textbf{B66}, 115205 (2002).

\bibitem{foot2} This
prediction is confirmed in PPV-DOO, where $E(S_{CT}) \sim 3.2$ eV
\cite{21} and $E(T_{CT}) \sim 3.1$ eV \cite{22}.

\bibitem{21} S. Frolov, Z. Bao, M. Wohlgennant, and Z. V. Vardeny,
\textit{ Phys. Rev.} \textbf{B65}, 205209 (2002).

\bibitem{22} A. P. Monkman, H. D. Burrows, L. J. Hartwell,
L. E. Horsburgh, I. Hamblett, and S.  Navaratnam,  \textit{Phys.
Rev. Lett.} \textbf{86}, 1358-1361 (2001).

\bibitem{24} See \textit{Charge and Energy Transfer Dynamics in Molecular
Systems}. V. May and O. Kuhn, Wiley-VCH, Berlin (2000).

\bibitem{foot10} In centro-symmetric light emitting polymers $n = 1$ and $j = 1$ corresponds to
the $1^1B_u$ state, and $n = 2$ and $j = 1$ corresponds to the
$m^1A_g$ state in the singlet sector, while in the
triplet sector
  $n = 1$ and $j = 1$ corresponds to the $1^3B_u$ state, and $n = 2$ and $j = 1$
 corresponds to the $1^3A_g$ state.

\bibitem{foot6} However, $H_{inter}$ can connect the inter-chain excitons to other even
parity intra-chain excitons, if the former lie higher in energy.
In that case the electron-hole recombination can be regarded as an
intra-chain processes, as the states must relax via the
intra-chain charge-transfer excitons, $S_{CT}$ and $T_{CT}$.

\bibitem{32} W. Barford, R. J. Bursill, and M. Yu Lavrentiev,  \textit{Phys. Rev.}
\textbf{B63}, 195108 (2001).

\bibitem{liess} M. Liess, S. Jeglinski, Z. V. Vardeny, M. Ozaki, K. Yoshino, Y. Ding, and T. Barton, \textit{Phys. Rev.} \textbf{B56}, 15712 (1997).

\bibitem{moore} E. E. Moore, W. Barford, and R. J. Bursill, \textit{Relaxation energies and excited state structures in poly(para-phenylene)}, submitted to Phys. Rev. B.

\bibitem{hertel} D. Hertel, S. Setayesh, H.-G. Nothofer, U. Scherf, K. M\"ullen and H. B\"assler, \textit{Advanced Materials}, \textbf{13}, 65 (2001).

\bibitem{frolov} S. Frolov, M. Liess, P. Lane, W. Gellermann, and Z. Vardeny, \textit{Phys. Rev. Lett.} \textbf{78}, 4285 (1997).

\bibitem{barford2004} W. Barford and E. E. Moore, \textit{An estimate of the inter-system crossing time in light-emitting polymers}, submitted to Phys. Rev. B.

\bibitem{33} D. Beljonne, A. Ye, Z. Shuai, and J. L. Br\'edas, \textit{Adv. Funct. Mater.}, \textbf{14}, 684 (2004)


\end{references}
\end{document}